\documentclass[twocolumn,amssymb,showpacs, amsmath,nobibnotes,nofootinbib, aps, prl]{revtex4}
\usepackage[dvips]{graphicx}
\begin{document}
\title{Anisotropic low field behavior and the observation of flux jumps in CeCoIn$_5$}
\author{S. Majumdar}
\email{phshd@warwick.ac.uk} 
\author{M. R. Lees}
\author{G. Balakrishnan}
\author{D. McK Paul}
\affiliation{Department of Physics, University of Warwick, Coventry CV4 
7AL, United Kingdom}
\begin{abstract}
The magnetic behavior of the heavy fermion superconductor CeCoIn$_5$ has been  investigated. The low field magnetization data show  flux jumps in the mixed state of the superconducting phase in a restricted range of temperature. These flux jumps begin to disappear below 1.7 K, and are completely absent at 1.5 K. The magnetization loops are asymmetric, suggesting that  surface and geometrical factors dominate the pinning in this system. The lower critical field ($H_{c1}$), obtained  from the magnetization data, shows a linear temperature dependence and is anisotropic. The calculated penetration depth ($\lambda$) is also anisotropic, which is consistent with the observation of an anisotropic superconducting gap in CeCoIn$_5$. The critical currents, determined from the high field isothermal magnetization loops, are comparatively low (around 4$\times$ 10$^3$ Acm$^{-2}$ at 1.6 K and 5 kOe).
\end{abstract}
\maketitle

Superconductivity in the heavy fermion compounds is unconventional in nature. Over the last two decades, several Ce and U based heavy fermion superconductors  have been discovered with superconducting transition temperatures ($T_c$) below 1 K. The magnetism and superconductivity are interrelated  in these compounds and it is argued that the superconducting state emerges out of the magnetic correlations rather than from any phonon mediated interactions~\cite{mathur}. Recently, a new class of heavy fermion compounds with the  general formula Ce$M$In$_5$ ($M$ = Co, Ir or Rh) has been discovered which show anomalous superconducting properties. These compounds have a quasi-two dimensional crystal structure consisting of CeIn$_3$ layers parallel to the $ab$ plane. CeIrIn$_5$ is superconducting below 0.4 K and CeRhIn$_5$ shows superconductivity only under applied hydrostatic pressure. CeCoIn$_5$ is a superconductor at ambient pressure with a $T_c$ = 2.3 K, which is relatively high compared to the  $T_c$'s of other heavy fermion superconductors. As a result, CeCoIn$_5$ provides us with a unique opportunity to investigate the nature of the superconductivity in this class of compounds.  Magnetization measurements~\cite{petrovich} indicate that the superconductivity in this layered compound is anisotropic in nature. The upper critical fields ($H_{c2}$) at 1.5 K  have been reported to be around 80 kOe and 30  kOe for magnetic fields applied parallel  and perpendicular to the $ab$ plane respectively~\cite{tayma}. Recent heat capacity and thermal conductivity measurements  indicate that the superconductivity in CeCoIn$_5$ is of  non-BCS character with anisotropic gap formation at the Fermi surface~\cite{heatcap,thermal}.

\par
The values of the upper and lower critical fields are important parameters that help characterize the nature of a superconductor. They enable us to estimate the microscopic superconducting length scales such as, the penetration depth $\lambda$ and the coherence length $\xi$. An exact determination of the critical fields, particularly $H_{c1}$, is often difficult due to demagnetization effects and the quality of the material available. However, it is possible to get convincing $H_{c1}$ data by careful measurements and analysis on a high quality single crystal sample. In this paper we report on a detailed magnetic investigation of CeCoIn$_5$ single crystals. We have obtained the temperature dependence of $H_{c1}$ for CeCoIn$_5$. The characteristic superconducting parameters $\lambda$ and $\xi$ were also calculated from the critical field values.  In addition, we have also investigated the critical current density of the material.

\par
Single crystals of CeCoIn$_5$  were prepared by the  indium flux  technique.  The crystals grew in the form of thin rectangular plates with an area of 2-3 mm$^2$ and a thickness of 0.2-0.3 mm. Concentrated hydrochloric acid was used to remove any  residual indium from the surface of the crystals.  X-ray Laue diffraction was performed to determine the crystallographic axes of the samples and  it was seen that the $ab$-planes of the crystals coincide with  the rectangular faces of the plates.  An Oxford Instruments vibrating sample magnetometer (VSM)  was used to measure the magnetization down to 1.5 K. The measurements were carried out on several crystals obtained from different batches. We have found no noticeable difference between  the magnetization behavior of these crystals. The data presented here are the results of measurements on a rectangular crystal (1.24 $\times$ 0.75 $\times$ 0.09 mm$^{3}$) where  the shortest dimension is along the $c$-axis.
 
\begin{figure}[t]
\vskip 0.4 cm
\centering
\includegraphics[width = 8.5 cm]{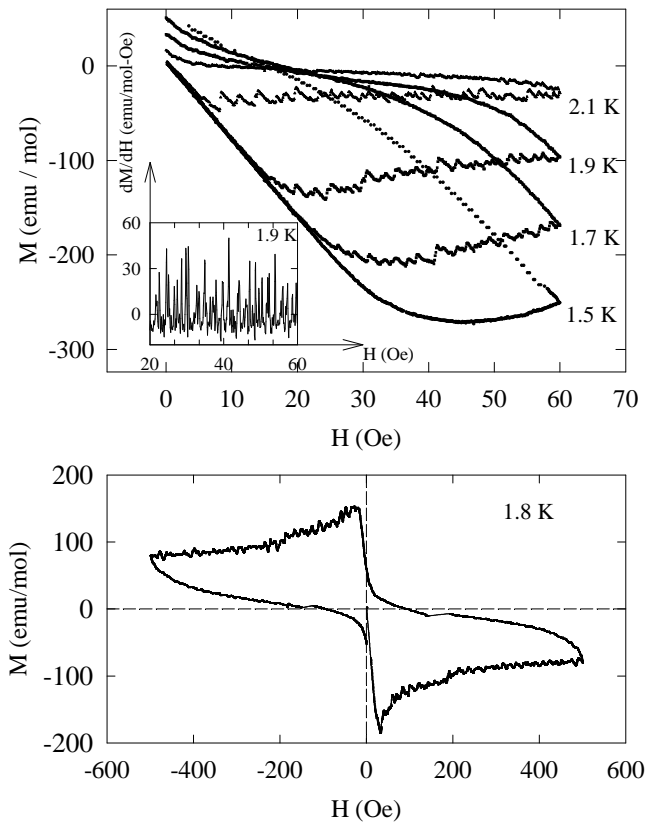}
\caption {The upper panel shows the low field (up to 50 Oe)  magnetization ($M$) versus  field ($H$)  data at different temperatures with the applied magnetic field parallel to the $a$ direction with a  field sweep rate of 5 Oe/minute. The inset shows the derivative of $M$ with respect to $H$ to depict the sharpness of  flux jumps. The lower panel shows a four-quadrant $M$ versus $H$ loop at 1.8 K in the same geometry as above up to $H$ = 500 Oe with a field sweep rate of 50 Oe/min.}
\label{fig1}
\end{figure}

The low field isothermal magnetization data of the compound CeCoIn$_5$ are shown in Fig.1. with the  field applied parallel to the $a$-axis  of the crystal. The data were collected at each  temperature after zero field cooling from 5 K.  In order to minimize the effect of residual flux trapped in the superconducting coil of the VSM, the magnet was degaussed before each measurement  by  applying a damped oscillatory field cycle. The sample chamber in the VSM was  flooded with helium exchange gas to ensure  temperature stability during the measurements.The magnetization loops below the superconducting transition temperature of 2.3 K show hysteresis  typical of a type II superconductor. 
\par
In order to calculate the lower critical field ($H_{c1}$), it is essential to take into account the demagnetization effect of the sample, because at  low field, the field correction is comparable to the applied magnetic field.  The demagnetization correction was performed using the relation $H_{eff} = H_{app} - 4\pi N_iM$, where $H_{app}$ is the applied external field (in Oe), $H_{eff}$ is the effective field (in Oe) on the sample after correction, $N_i$ is the demagnetization factors for different directions ($i = a, b, c$) and $M$ is the magnetization (emu/cm$^3$) of the sample. Assuming that the sample is ellipsoidal in shape, the values of $N_i$,  obtained from  reference~\cite{demag}, are found to be 0.03, 0.05 and 0.92 for  the $a$, $b$, and $c$ directions respectively. 

\begin{figure}[t]
\centering
\includegraphics[width = 8.5 cm]{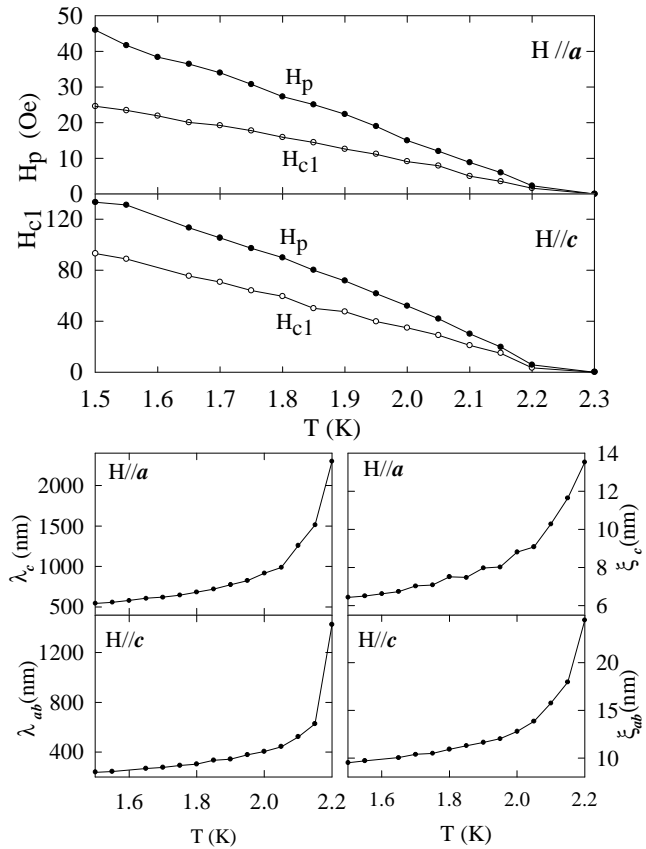}
\caption {The upper panels show the temperature dependence of the lower critical field ($H_{c1}$) and the full penetration field ($H_{p}$) for CeCoIn$_5$ with the applied field parallel  to the $a$ and $c$ directions. The superconducting penetration depth ($\lambda$) and the correlation length ($\xi$) are plotted as a function of temperature in the lower panels.}
\label{fig2}
\end{figure}

The lower critical field of a superconductor is defined as the onset of the deviation from an ideal diamagnetic ($M/H = -1/4\pi$) behavior. For the calculation of the lower critical field, we have subtracted the ideal  linear diamagnetic response ($M_{dia}(H) = -H/4\pi$ ) from our magnetization data,  to obtain the deviation $\delta M = M - M_{dia}$). This deviation $\delta M$ varies as ($H - H_{c1})^2$ around $H_{c1}$~\cite{hc1}. Thus the value of $H_{c1}$ can be obtained from the $H$ intercept of a $(\delta M)^2$ versus $H$ plot. We have also calculated the full penetration field, $H_p$ at different temperatures, by noting the magnetic field at which maximum diamagnetic signal is observed in the $M$ versus $H$ measurements. For the calculation of $H_{c1}$ and $H_p$, we have used the demagnetization corrected field $H_{eff}$. 

\par
Fig. 2 shows the variation of the lower critical field and the full penetration field with temperature for the field applied parallel to the $a$ and $c$ directions respectively. The $H_{c1}$ and $H_{p}$ values fall  almost linearly with temperature  for both  these directions. This  linear behavior of $H_{c1}$ is unusual in low $T_c$ superconductors.  The $H_{c1}$ values are anisotropic in magnitude with respect to the $a$ and $c$ directions. This is not unexpected given that CeCoIn$_5$ is a layered compound and that it has shown anisotropies in its resistive and magnetic behavior~\cite{petrovich}.

\par
The $M$-$H$ loops are found to be  asymmetric (see Fig. 1) with respect to the $M$ = 0 axis. This asymmetry  is also seen in the magnetization data taken up to 500 Oe. This  observation suggests that the magnetization behavior at low fields is dominated by surface and geometrical barriers rather than bulk pinning. We have scaled the low field magnetization data by the full penetration field magnetization ($M_p$, the magnetization at $H_p$) and shown that at low fields ($H \leq$ 100 Oe) the plots (not shown) of $M$/$M_p$ versus $H$/$H_p$ for different temperatures collapse on to  each other. This  confirms that  temperature independent pinning  mechanisms, such as  surface and geometrical barrier effects, are present in this material. 
\par
In common with other magnetic superconductors ({\it e.g.} rare earth borocarbides, UPt$_3$), CeCoIn$_5$ has a field dependent positive contribution to the  magnetization ($M_{para}$) superimposed on top of  the diamagnetic response. As a result, the magnetization becomes positive well below the upper critical field. It is often difficult to discern the true nature of the hysteresis loop at high fields, where $M_{para}$ is large. Nevertheless, we have  estimated  $M_{para}$ from the magnetization data (at 2.4 K) just above the $T_c$ of CeCoIn$_5$. We have then subtracted this $M_{para}$ from the magnetization data obtained below $T_c$, assuming that the field dependence of  $M_{para}$ remains unchanged within the temperature range 1.6-2.4 K~\cite{mpara}. The resultant loops ($M-M_{para}$ versus $H$) are also asymmetric with respect to the $M$ = 0 line, indicating that surface and geometrical effects are important even in the high field state.

\par
Another interesting observation in the low field measurements is the  flux jumps in the magnetization data. The observed jumps are irregular and non-periodic (see Fig. 1). The flux jumps are observed for measurements made in {\it increasing} field with the magnetic field applied along either the $a$ or the  $c$ axes. The jumps are completely absent (for $H \parallel a$ axis) or very weak (for $H \parallel c$) when the field is ramped {\it down}. The magnitude  of these jumps is largest  at  2.1 K  and  then decreases slowly below 1.7 K. The flux jumps are completely absent at 1.5 K (see Fig. 1). The magnitude of the largest jump at 2.1 K is 0.16 emu/c.c. This corresponds to an entry of $\sim$ 6000 flux quanta ($\Phi_0$) into the sample. Flux jumps are observed in many type II superconductors~\cite{fluxjump}. During the field sweep, a  small perturbation in the flux distribution in the critical state can give rise to a temperature fluctuation, which can in turn  result in  the movement of a flux bundle within  the sample. A jump is then observed in the magnetization data. The gradual disappearance of these flux jumps below 1.7 K indicates that there is some difference in the nature of the flux distribution (due the variation of pinning mechanism or the thermal diffusibility) below  1.7 K. Note however, that heat capacity and  thermal  conductivity  data for CeCoIn$_5$ contain no unusual features below $T_c$.   

\par
In order to understand the nature of the flux jumps, we have measured the magnetization data with  different field sweep rates (5 Oe/min to 10 kOe/min). In all cases, there are a large number of flux jumps that are small in magnitude. No avalanche-like large flux jumps, (which can drive the system above $T_c$), were observed even at the highest sweep rate. The flux jumps were also observed in CeCoIn$_5$ crystals of different shape and size and  the qualitative features of the flux jumps were completely reproducible. It appears that the observed flux instabilities are due to local flux entry through the surface or geometrical barriers rather than any global instability. The asymmetry of the flux jumps with respect to the increasing and decreasing part of the magnetization loop clearly indicate the existence of  barriers at the surface (such as those of the Bean-Livingstone type) which prevent the smooth entry of flux lines into the sample, but are ineffective during  flux expulsion. 
\par
Within the Ginzburg-Landau approximation, the characteristic superconducting length scales $\xi$ and $\lambda$ can be estimated from the knowledge of $H_{c1}$ and $H_{c2}$ using the following relations:
\begin{eqnarray}
\label{kappa1}
H_{c1} = \frac{ \Phi_0 \ln \kappa}{4\pi \lambda^2}, ~H_{c2} = \frac{\Phi_0}{2\pi \xi^2} \nonumber \\
H_{c2}/H_{c1} = 2\kappa^2/\ln \kappa, \kappa = \lambda/\xi 
\end{eqnarray}

The values of $H_{c2}$ were obtained from the high field $M$ versus $H$ data (not shown here) at different temperatures for  fields applied parallel to the $a$ and the $c$ axes. The field values where the irreversibility between the increasing and the decreasing branches disappears were taken as the upper critical field of the sample. Our values match well with the previously reported values of $H_{c2}$ from magnetization data~\cite{tayma}. Using equation~\ref{kappa1} we have obtained the values of $\kappa$ (= $\lambda/\xi$).  At 1.6 K, these are about 90 and 25 for the field applied parallel to the $a$ and $c$ axes respectively. Fig. 2 shows the temperature dependence of $\lambda$ and $\xi$ for both the directions ($H \parallel a$ and $H \parallel c$). Both parameters are  anisotropic. The ratios $\lambda_{\bf c}/\lambda_{\bf ab}$  and  $\xi_{\bf ab}/\xi_{\bf c}$  are $\sim$ 2.3 and 1.5 in the temperature range 1.5 to 2.1 K,  where the subscripts $ab$ and $c$  denote the $ab$ plane and $c$ axis respectively. Since the penetration depth is directly proportional to the square root of the effective mass ($m^*$), this implies that there is a large anisotropy in the effective mass within the material ($\Gamma = m_c^*/m_{ab}^* \sim 5.3$)\cite{fnote}. The anisotropy  of $\lambda$ is clearly consistent with the observation of nodes in the superconducting gap at particular points of the Fermi surface~\cite{thermal}. Anisotropies in $H_{c1}$ and $\lambda$ are  also observed in  other heavy fermion superconductors with anisotropic superconducting gap  like UPt$_3$~\cite{upt3}. Our calculated values of $\lambda$ from the magnetization measurements (for $H \parallel a$) are qualitatively similar and quantitatively close ($\lambda \sim$ 425 nm in reference~\cite{tunneling} as compared to our value of 540 nm at 1.5 K)  to  the values obtained recently from tunneling experiments~\cite{tunneling}. The reasonably large value of $\lambda$ ($\sim$ 500 nm)  observed for CeCoIn$_5$ is typical of the heavy fermion superconductor ($\lambda \sim$ 1000 nm for UPt$_3$~\cite{upt3}). 

\begin{figure}
\centering
\includegraphics[width = 8.8 cm]{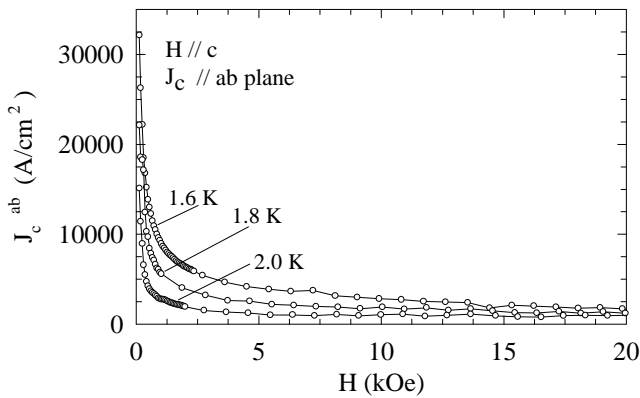}
\caption {Critical current density in the $ab$ plane  at different temperatures for CeCoIn$_5$ plotted against the applied magnetic field.}
\end{figure}

The critical current density, $J_c$, is not an intrinsic parameter of a superconductor. Furthermore, in  systems such as CeCoIn$_5$, where  surface effects are present, one should be careful when considering  the critical currents within the material. Nevertheless, the use of a Bean-like critical state model can provide us with an indication of the strength  of the pinning within the system. For a thin rectangular plate-like superconducting sample (sides $t$ and $\ell$, $\ell > t$) with an applied magnetic field perpendicular to the plane of the plate, $J_c$ on the surface of the plate is given by~\cite{jceqn}:
\begin{equation}
J_c = 20(M\downarrow - M\uparrow)[t(1 - t/3\ell)]^{-1}
\label{jc}
\end{equation}
where $M\downarrow$  and  $M\uparrow$ are the magnetization (in gauss) for the decreasing and increasing fields respectively. This relation is valid only when we have  isotropic critical currents perpendicular to the applied field. Since the superconducting properties of  CeCoIn$_5$ appear to be  isotropic in the $ab$ plane, it is possible to apply equation~\ref{jc} in order to calculate $J_{c}^{ab}$ the critical current density in the $ab$ plane. Figure 3 shows the variation of $J_{c}^{ab}$ as a function of the applied field. The $J_c$ values are derived from the high field magnetization data measured with the field parallel to the $c$ direction. From figure 3 it is clear that $J_c^{ab}$ drops smoothly with  increasing magnetic field and temperature. No unusual variation in the behavior of $J_c^{ab}$ with field or temperature was observed. The value of $J_c^{ab}$ is about 4$\times$ 10$^3$ Acm$^{-2}$ at 1.6 K  in an applied field of 5 kOe, which is a few orders of magnitude lower than  some high $T_c$ materials~\cite{jceqn}. However, a low value of $J_c$ is not unusual when the bulk pinning is weak. For example, a critical current of similar magnitude ($\sim$ 10$^3$ A/cm$^2$ at $T/T_c$ = 0.7 with 5 kOe of applied field) has been observed in YNi$_2$B$_2$C crystals~\cite{yni2b2c}, where the surface and geometrical effects are predominant over the bulk pinning. 

\par
CeCoIn$_5$ has a layered structure consisting of quasi-two dimensional CeIn$_3$ building blocks parallel to the a-b plane.  The observed anisotropies in $H_{c1}$, $\lambda$ and $\xi$, clearly support the idea that the  CeIn$_3$ layers have an important influence on the superconducting properties of  CeCoIn$_5$. The observation of flux jumps is interesting,  however it is not clear at present why the flux jumps disappear below 1.7 K. Our measurement of the critical currents  show no unusal change below 1.7 K. 
\par
We thank Dr. C. D. Dewhurst(ILL, Grenoble) for useful discussions and acknowledge the support of the EPSRC(UK) for this project.


\begin{thebibliography}{99}

\bibitem[1]{mathur} N. D. Mathur, F. M. Grosche, S. R. Julian, I. R. Walker, D. M. Freye, R. K. W. Haselwimmer, and  G. G. Lonzarich Nature (London) {\bf 394}, 39 (1998).

\bibitem[2]{petrovich} C. Petrovich P. G. Pagliuso, M. F. Hundley, R. Movshovich, J. L. Sarrao, J. D. Thompson, Z. Fisk, and P. Monthoux, J. Phys. Condens. Matter {\bf 13}, L337 (2001).
 
\bibitem[3]{tayma} T. Tayama, A. Harita, T. Sakakibara, Y. Haga, H. Shishido, R. Settai, and Y. Onuki, Phys. Rev. B {\bf 65}, 180504 (2002). T. P. Murphy, Donavan Hall, E. C. Palm, S. W. Tozer, C. Petrovic, and Z. Fisk, R. G. Goodrich, P. G. Pagliuso, J. L. Sarrao, and J. D. Thompson, Phys. Rev. B {\bf 65}, 100514 (2002). 


\bibitem[4]{heatcap} R. Movshovich , M. Jaime , J.D.  Thompson , C. Petrovic , Z. Fisk , P.G. Pagliuso , J.L.  Sarrao, Phys. Rev. Lett. {\bf 86}, 5152 (2001).
 
\bibitem[5]{thermal} K. Izawa, H. Yamaguchi, Yuji Matsuda, H. Shishido, R. Settai, and Y. Onuki, Phys. Rev. Lett. {\bf 87}, 57002 (2001).

\bibitem[6]{demag} J. A. Osborn, Phys. Rev. {\bf 67}, 351 (1945).

\bibitem[7]{hc1} V. V. Metlushko and G. G\"{u}ntherodt, V. V. Moshchalkov and Y. Bruynseraede, and M. M. Lukina, Phys. Rev. B {\bf 47}, 8212 (1993).

\bibitem[8]{mpara} An extrapolation of the normal state (2.3 to 5 K) $M$ versus $H$ data for  CeCoIn$_5$ collected in a field of 20 kOe  down to 1.5 K indicates only a 2\% increase  in $M_{para}$ between 2.4 and 1.6 K.  

\bibitem[9]{fluxjump} M.E. McHenry, H. S. Lessure, M. P. Maley, J. Y. Coulter, I. Tanaka,  and H. Kojima, Physica C {\bf 190}, 403 (1992) and references therein.

\bibitem[10]{tunneling} S. \"{O}zcan, D. M. Broun, B. Morgan, R. K. W. Haselwimmer, J. L. Sarrao, Saied Kamal, C. P. Bidinosti, P. J. Turner, M. Raudsepp and J. R. Waldram., cond-mat/0206069 (2002).

\bibitem[11]{jceqn} E. M. Gyorgy, R. B. van Dover, K. A Jackson, L. F. Schneemeyer, and J. V. Waszczak, Appl. Phys. Lett. {\bf 53}, 283 (1989).

\bibitem[12]{fnote} This is an estimate, since the ratios $\lambda_{\bf c}/\lambda_{\bf ab}$  and  $\xi_{\bf ab}/\xi_{\bf c}$ differ.

\bibitem[13]{upt3}Zuyu Zhao, F. Behroozi, S. Adenwalla, Y. Guan, and J. B. Ketterson, Bimal K. Sarma, and D. G. Hinks, Phys. Rev. B, {\bf 43}, 13720 (1991). C. Broholm, G. Aeppli, R. N. Kleiman, D. R. Harshman, D. J. Bishop, and E. Bucher, D. Ll. Williams, E. J. Ansaldo, R. H. Heffner, Phys. Rev. Lett. {\bf 65}, 2062 (1990).
 
\bibitem[14]{yni2b2c} S. S. James, C. D. Dewhurst, R. A. Doyle, D. McK Paul, Y. Paltiel, E. Zeldov and A. M. Cambell, Physica C {\bf 332}, 173 (2000).


\end{thebibliography}
\end{document}